\documentclass[12pt]{article}

\usepackage{amsmath}
\usepackage{graphicx}% Include figure files
\usepackage{color}
\usepackage[compat=1.0.0]{tikz-feynman}
\usepackage{feynmp-auto}
\usepackage[colorlinks=true, urlcolor=blue, linkcolor=blue, 
citecolor=blue]{hyperref}
\usepackage{cite}

\def \lsim{\mathrel{\vcenter
     {\hbox{$<$}\nointerlineskip\hbox{$\sim$}}}}

\newcommand{\dd}{{\rm d}}
\newcommand{\N}{{\cal N}}

\newcommand{\beq}{\begin{equation}}
\newcommand{\eeq}{\end{equation}}
\newcommand{\beqa}{\begin{eqnarray}}
\newcommand{\eeqa}{\end{eqnarray}}
\newcommand{\beqar}{\begin{eqnarray*}}
\newcommand{\eeqar}{\end{eqnarray*}}

\begin{tiny}

\end{tiny} %{\vskip-2ex$_{#1}%$\label{#1}}

\begin{document}
\thispagestyle{empty}
$\,$

\vspace{32pt}

\begin{center}

\textbf{\Large Monochromatic neutrinos from dark matter\\}
\textbf{\Large through the Higgs portal}

\vspace{50pt}
Pablo de la Torre, Miguel Guti\'errez, Manuel Masip
\vspace{16pt}

\textit{CAFPE and Departamento de F{\'\i}sica Te\'orica y del Cosmos}\\
\textit{Universidad de Granada, E-18071 Granada, Spain}\\
\vspace{16pt}

\texttt{pdelatorre,mgg,masip@ugr.es}

\end{center}

\vspace{30pt}

\date{\today}% It is always \today, today,

\begin{abstract}

We define a minimal
model of dark matter with a fermion singlet
$\chi$ coupled to the visible sector through the Higgs portal and with a heavy
Dirac neutrino $N$ that opens the annihilation channel $\chi \chi \to N \nu$.
The model provides the observed relic abundance consistently with bounds from direct searches and
implies a monochromatic neutrino signal at 10 GeV--1 TeV in indirect searches. In particular, 
we obtain the capture rate of $\chi$ by the Sun and show that the signal could be above the {\it neutrino floor} produced
by cosmic rays showering in the solar surface. In most
benchmark models this solar astrophysical background is above the expected dark matter signal, so the model that we propose
is a canonical example of  WIMP not  excluded by direct searches that
could be studied at neutrino telescopes and also at colliders.

\end{abstract}

\vfill
\eject

\section{Introduction} 

The WIMP paradigm provides arguably the simplest scenario for the dark matter (DM) of the universe. It represents a {\it smooth} extension of the Standard Model (SM), involving a physics that is new but that sounds quite  familiar, not too exotic. The  WIMP $\chi$ that constitutes the DM is stable just like the proton, with a matter parity playing the  role of the baryon number. Its mass $m_\chi$ may introduce a new scale in physics, but this scale 
should not be far from the electroweak (EW) one. The same occurs with the mediators
of its interactions with the visible sector: the EW gauge bosons do not work, but the Higgs boson may \cite{Lebedev:2021xey,Arcadi:2021mag} and, in any case, the interaction required should define EW-like cross sections. In the early universe the WIMP is in thermal equilibrium with the SM particles, until it decouples and its abundance freezes out just like happens to neutrinos, nothing extraordinary. This simplicity plus the fact that the scenario has interesting implications at direct, indirect and collider searches has made the WIMP the most popular DM candidate for over 40 years \cite{Turner:1983anb}. 

Indeed, WIMPs like the lightest odd particle in SUSY models with R-parity, in Little Higgs models with T-parity or in models with compact dimensions and KK-parity, are part of a  setup that completes the SM at the TeV scale. The main motivation for these models was to solve the hierarchy problem, being the presence of a suitable DM candidate an interesting but additional feature. However, the LHC has basically 
discarded {\it naturality} as a guiding principle in the search for non-standard TeV physics. At this point it may be more effective an approach based on {\it minimality}, where one assumes that the new physics may appear with equal likelihood wherever it 
is not experimentally excluded. 
It is not that SUSY can not explain the DM, an anomaly in the muon $g-2$ or a peak followed by a dip in di-Higgs production at the LHC, it is that there is no need nor a strong motivation for the whole setup: keep just the few elements that are enough to explain the effect and save you the hustle of hiding the rest of the setup. 

In that context, we will consider here a very minimal model of WIMP. The first basic question that such model should answer is whether it can provide the relic abundance $\Omega_c h^2\approx 0.12$ that we observe or, more precisely, if it can do it consistently with the bounds from direct searches. These searches push the WIMP-nucleon elastic cross section down, which tends to reduce the annihilation cross section and imply too large values of $\Omega_c$. Second, it is interesting to know whether the model can be probed in indirect searches once the bounds from direct searches are imposed. Moreover, if these bounds reached the neutrino floor where the DM signal faces an (almost) irreducible background \cite{OHare:2021utq}, could we still expect any positive results in indirect searches? We will focus on the signal from DM annihilation in the Sun at neutrino telescopes. On one hand, the same collisions with solar nuclei that capture the WIMP are also probed in direct searches. On the other hand, high energy cosmic rays (CRs) showering in the solar surface produce a flux of neutrinos \cite{Seckel:1991ffa,Edsjo:2017kjk,Masip:2017gvw,Gutierrez:2019fna}  that, although not observed yet, represents a neutrino floor analogous to the one in direct search experiments \cite{Gutierrez:2022mor}.

The DM of the universe may involve physics at basically any scale. The thermal WIMP at $m_\chi=10$ GeV--$10$ TeV is just one of the many possibilities, but it is the only one that can be probed in direct, indirect and collider searches. Here we define a minimal setup that is consistent with the data in a wide range of parameters and that could imply an observable signal at  neutrino telescopes and also at colliders. The model is a variation of the Higgs portal proposed a couple of decades ago \cite{Kim:2006af,Lopez-Honorez:2012tov,Baek:2011aa,Djouadi:2011aa} extended with a heavy Dirac neutrino. Due to its simplicity, it may help us to calibrate how constrained the WIMP paradigm currently is and what may be the chances to get a signal in future searches.

\section{The model} 

Let us use 2-component spinors of left handed chirality to define the model.\footnote{In this 2-component notation $e_\alpha$ and ${e^c}_{\!\!\alpha}$ are the electron and the positron both {\it left}, whereas their conjugate-contravariant ${\bar e}^{\,\dot\alpha}$ and ${\bar {e^c}}\,^{\!\dot\alpha}$ are {\it right} spinors. The 4-component electron in the chiral representation of $\gamma^\mu$ is then 
$\Psi_e=\begin{pmatrix} e_\alpha \\ {\bar {e^c}}\,^{\!\dot\alpha} \end{pmatrix}$ [with $\bar \Psi_e=\left(  {e^c}\,^{\!\alpha}\; \bar e_{\dot \alpha} \right)$], while 
 $\Psi_\chi=\begin{pmatrix} \chi_\alpha \\ {\bar {\chi}}\,^{\!\dot\alpha}\end{pmatrix}$ is a Majorana fermion.}
We take $\chi$ as a Majorana singlet, although the results would be similar if the DM particle is a Dirac fermion. In addition, we introduce a Dirac singlet $(N,N^c)$ of similar mass $m_N$; these fields are needed to make the SM neutrinos massive through an inverse see-saw (we assume that the rest of them are heavier and/or less coupled to the active neutrinos).  We then assign odd matter parity to $\chi$ and $+1$ ($-1$) lepton number to $N$ ($N^c$). If we consider an effective 
theory valid at energies below $\Lambda$, the relevant part of the Lagrangian is just
\beq
-{\cal L} \supset {1\over 2}\, m_\chi \, \chi \chi + {c_s\over \Lambda} \,H^\dagger H\, \chi \chi
+i \,{c_a\over \Lambda} \,H^\dagger H\, \chi \chi +\; m_N \, N N^c \nonumber 
\label{lag}
\eeq
\beq
\hspace{0.5cm} +\, y_N\, H L N^c+i\, {c_N\over \Lambda^2} \left( N N^c + \bar N \bar N^c \right) \left( \chi \chi \right) + {\rm h.c.}\,,
\eeq
where $H=(h^+\; h^0)$ is the SM Higgs, $L=(\nu \; \ell)$ a lepton doublet assumed mostly along the $\tau$ flavor, and the six parameters defining the model are real. Other possible terms, like the dim-5 operator $H^\dagger H\, N N^c$, give subleading effects
that we will comment later on. 
Notice 
that in 4-spinor notation (see the footnote) the two operators connecting $\chi$ with $H$ plus their conjugate are just $(c_s/\Lambda)  \,H^\dagger H\, \bar \Psi_\chi  \Psi_\chi-(i c_a/\Lambda)  \,H^\dagger H\, \bar \Psi_\chi \gamma_5 \Psi_\chi$, whereas the dim-6 operator would read
$(-ic_N/\Lambda^2)  \, ( \bar \Psi_N  \Psi_N ) ( \bar \Psi_\chi \gamma_5 \Psi_\chi ) $.
These operators may result after integrating out heavy particles, in particular, 
\begin{itemize}
\item 
A real scalar singlet $s$ even under the matter parity coupled to the Higgs through the trilinear $s H^\dagger H$, to $\chi$ through the Yukawa $s \chi\chi$ (the real and imaginary parts of this coupling would contribute, respectively, to $c_s$ and $c_a$) and to the heavy neutrino through $s N N^c$. 
\item
A vectorlike lepton doublet $(D, D^c)$ odd under the matter parity coupled to $H D \chi$ and $H^\dagger D^c \chi$ may generate values of $c_s$ and $c_a$ that will depend on the relative complex phase in these two Yukawas. 
\end{itemize}
Notice also that in the first case the SM Higgs will necessarily mix with $s$ and the mass eigenstate will get a small singlet component, whereas in the second model it is the fermion singlet $\chi$ who will mix with the neutral fermions in $(D,D^c)$ and acquire a small doublet component. This could introduce interesting effects in the UV complete model, but they are subleading in the regime that we assume with $s$ and $(D,D^c)$ heavier than $\chi$ and $N$.

Let us then continue with the effective model.  At the EW minimum, $\langle H\rangle=(0\; v/\sqrt{2})$, 
in the unitary gauge  the Lagrangian includes (we use primes to denote mass eigenstates)
\beq
-{\cal L}\supset {1\over 2}\, m'_\chi \, \chi \chi  + m'_N \, N' N^c + {\tilde y_N\over \sqrt{2}} \, h\, \nu' N^c 
+  {c_s+i c_a\over 2\Lambda} \left( hh\, \chi \chi +2h\, \chi \chi \right) 
 \nonumber
\eeq
\vspace{-0.5cm}
\beq
\hspace{1cm}  +\, 
{i c_N\over \Lambda^2} \, ( c_\alpha \, N' N^c+ s_\alpha \, \nu'  N^c + c_\alpha\,  \bar N' \bar N^c+ s_\alpha\,  \bar \nu'  \bar N^c) \, (\chi \chi)
 + {\rm h.c.}\,,
\eeq
where $N'=c_\alpha\, N + s_\alpha \,\nu$ and $N^c$ combine into a heavy Dirac neutrino of mass
\beq
m'_N=\sqrt{\left({y_Nv\over \sqrt{2}}\right)^2 +  m_N^2}
\eeq
while the orthogonal combination $\nu'=c_\alpha\, \nu - s_\alpha \,N$ remains massless. The active-sterile mixing is just
$s_\alpha={y_N v\over \sqrt{2} \,m'_N}$, whereas the Yukawa coupling $\tilde y_N$, 
\beq
  {\tilde y_N}= {y_N}\, c_\alpha = 
  {\sqrt{2}\, m'_N\over v}\, c_\alpha \, s_\alpha
\eeq
may receive a contribution from the dim-5
operator $H^\dagger H\, N N^c$ that could be sizeable for $m'_N\lsim v$.
The mass of $\nu'$ and of the other two
SM neutrinos should then be implemented through an inverse see-saw mechanism. A particular model along these lines has been recently proposed as a solution to the so 
called $H_0$ tension. \cite{Cuesta:2021kca,Cuesta:2023awo}.

In summary, the setup includes 6 parameters (we drop all the primes): 
the mass $m_\chi$ of the  DM particle; the two couplings $c_{s,a}/\Lambda$ connecting $\chi$ with the Higgs; 
the mass $m_N$ of a heavy Dirac 
neutrino ($N,N^c$); the heavy-light Yukawa $\tilde y_N$ (correlated with the mixing $s_\alpha$), and the coupling $c_N/\Lambda^2$ connecting $\chi$ with the heavy neutrino.

\section{Cross sections} 

\begin{figure}[]
\begin{center}
\includegraphics[scale=0.75]{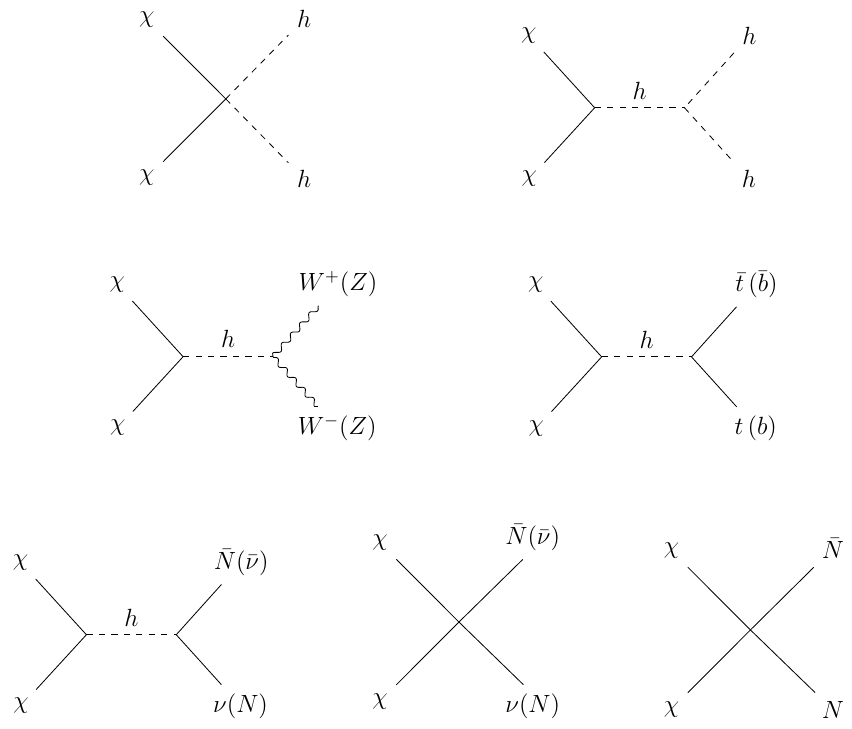}
\end{center}
\vspace{-0.5cm}
\caption{Leading diagrams in DM annihilation. 
\label{f1}}
\end{figure}

We are interested in the production of monochromatic neutrinos through $\chi \chi \to N \nu$ (we use $N\nu$ to indicate $\bar N \nu + N \bar \nu$, see the two relevant diagrams in Fig.~\ref{f1}), so we
will consider relatively large values of the coupling $\tilde y_N$ and of the four fermion coefficient $c_N$. 
In particular, notice that the Yukawa $\tilde y_N$ does not imply a mass for the active neutrino, it is
just constrained by collider bounds on the heavy-light mixing (assumed mostly along the tau flavor), $s_\alpha\le 0.1$ \cite{Fernandez-Martinez:2016lgt,Hernandez-Tome:2019lkb,Hernandez-Tome:2020lmh}. Since $s_\alpha = m_N/(174\;{\rm GeV})$, 
only values of $m_N$ above 1 TeV will allow for top-like Yukawa couplings. In any case, to avoid LEP bounds from 
 $Z\to N \nu$ plus $N\to \tau W^*,\nu Z^*$ we will not consider values of $m_N$ below $M_Z$. In turn, if the annihilation channel $\chi \chi \to N \nu$ is open this implies that $m_\chi>M_Z/2$. Throughout the analysis we will then take $m_\chi\ge 50$ GeV, $m_N\ge 100$ GeV, $s_\alpha= 0.1$ and $c_i/\Lambda < 1/(2m\chi),1/v$.

The DM annihilation in the early universe and in astrophysical environments will go through the diagrams in Fig.~\ref{f1}. If we neglect $c_N$ (the last two diagrams in the figure) 
the cross section can be written  
\beq
\sigma_{\rm ann}^{(1)} =  {c_s^2 \left( 1-{4m_\chi^2\over s} \right) + c_a^2\over 4 \pi \Lambda^2 \left(1-{4m_\chi^2 \over s}\right)^{1/2}} \,f(m_\chi)\approx 
 {c_s^2 \,\beta^2 + c_a^2 \over 4\pi \Lambda^2\, \beta} \,f(m_\chi)\,,
\label{ann}
\eeq
where $\beta$ is the velocity of $\chi$ in the center of mass frame and $f(m_\chi)=\sum_i f_i$ gives the contribution of the  
different channels for a given value of $m_\chi$. We obtain
\beqa
f_{hh}&=&\left( 1 + {3m_h^2\over s-m_h^2} \right) \sqrt{1-{4m_h^2 \over s}} \nonumber \\
f_{QQ}&=& {3m_Q^2 \left(s-4m_Q^2\right) \over \left(s-m_h^2\right)^2} \,\sqrt{1-{4m_Q^2 \over s}} \nonumber \\
f_{VV}&=& \frac{ 2m_{V}^{4} }{ \left(s-m_{h} \right)^{2} } \, \left[ 2 + \left(  1- \frac{s}{2m_{V}^{2}}  \right)^2 \right] \sqrt{1-{4m_V^2 \over s}} \nonumber \\
f_{N\nu}^{(1)}&=& {\tilde y_N^2 v^2 \left( s-m_N^2 \right)\over 2\left( s-m_h^2\right)^2 }  \left( {1-{m_N^2 \over s} }\right) 
\eeqa
with $Q=t,b$ and $V=Z,W$. If $c_N$ is sizeable we have to add
\beq
\sigma_{\rm ann}^{(2)}  = {c_N^2 \over 4\pi \Lambda^2 \left(1-{4m_\chi^2 \over s}\right)^{1/2}} \left(
f_{NN}+ f_{N\nu}^{(2)} \right)\,,
\label{ann2}
\eeq
where
\beqa
f_{NN}&=&   {c_\alpha^2 \left(s-4m_N^2\right) \over \Lambda^2}\,\sqrt{1-{4m_N^2 \over s}} \nonumber \\
f_{N\nu}^{(2)}&=& \left( {s_\alpha^2 \over \Lambda^2}-
 { \sqrt{2} \,c_a \, \tilde y_N v \over c_N \Lambda \left( s-m_h^2\right) } \right)
\left(  {s-m_N^2}\right) \left( {1-{m_N^2 \over s} }\right) .
\eeqa
Notice that at $s\approx  4m_\chi^2$ the channel  
$\chi \chi\to N \nu$ produces an active  neutrino of energy 
\begin{figure}[]
\begin{center}
\includegraphics[scale=0.4]{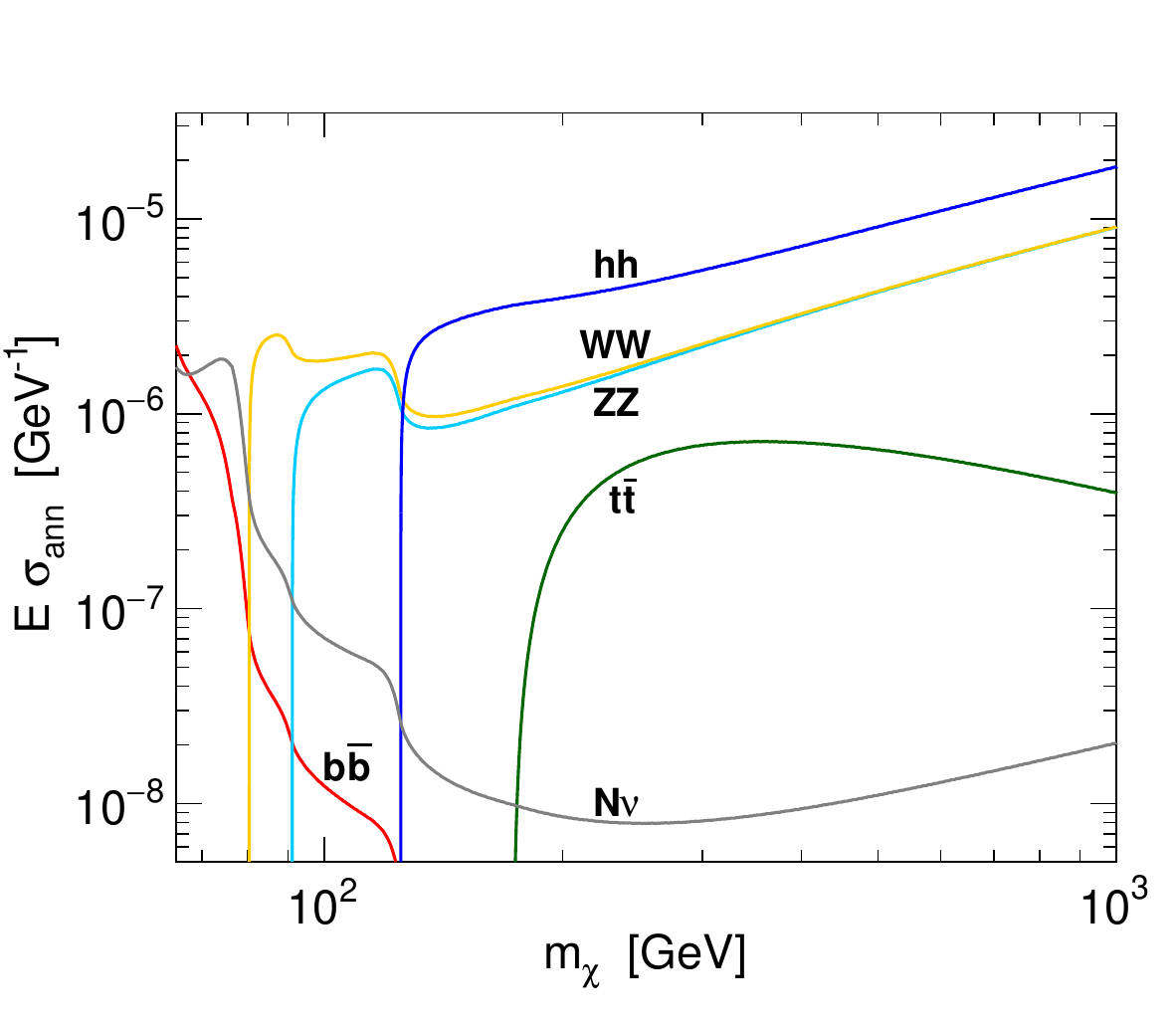}
\includegraphics[scale=0.4]{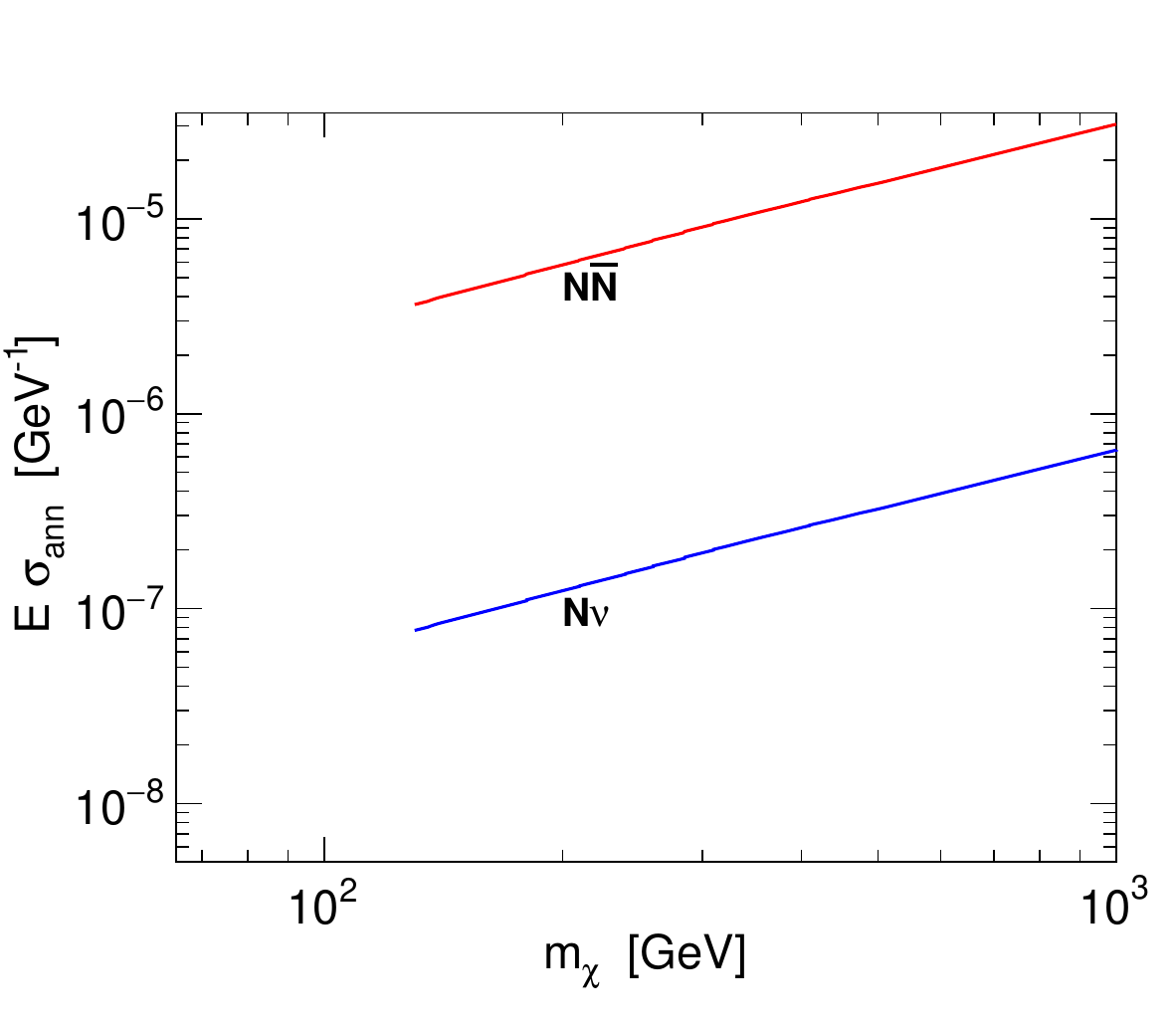}
\end{center}
\vspace{-0.5cm}
\caption{Different contributions to $\sigma_{\rm ann}$ for $c_N\to 0$, $m_N=1.4 \,m_\chi$ (left) and $c_{a} \to 0$, $m_N=0.7 \, m_\chi$  (right). We have taken $s_\alpha=0.1$ and $\beta=1/20$.
\label{f2}}
\end{figure}
\beq
E_\nu=m_\chi \left( 1-{m_N^2\over 4 m_\chi^2} \right).
\eeq
In Fig.~\ref{f2} we plot the contribution of the different channels for values of $m_\chi$ between 65 GeV and 1 TeV  in the 
limits $c_N\to 0$ and $c_{a} \to 0$.
Whenever 
$m_N<m_h$ and the Higgs  can decay $h\to N \nu$, we use the value of $y_N$ implying a 10\% branching ratio for this  channel (see discussion in section 5.1). In all the cases we have $m_N>M_{W,Z}$.

The second relevant cross section is the one describing the elastic scattering of $\chi$ 
with a nucleon $\cal N$, which is mediated by a Higgs in the $t$ channel. The Higgs boson 
couples to the quarks and (through heavy quark loops) to the gluons in the nucleon; at low energies
this induces a Higgs-nucleon Yukawa coupling
$g_{h\N}$ that is usually parametrized as $g_{h \N }=f_\N m_\N / v$ \cite{Cline:2013gha,Casas:2017jjg}, with $m_\N=0.94$ GeV and
$f_\N=0.30$ \cite{Alarcon:2011zs,Abdel-Rehim:2016won}. We obtain
\beq
\sigma(\chi \N\to \chi \N )
={4\over \pi}\, {c_s^2 + c_a^2 \,\beta^2 \over \Lambda^2} \left( {\mu_\N \,m_\N f_{\N} \over m_{h}^{2}} \right)^2\,.
\label{elastic}
\eeq
with $\mu_\N=m_\N m_\chi/(m_\N+m_\chi)$.

\section{Direct searches and relic abundance} 

Direct searches expect to see the recoil of a nucleus hit by a DM particle of velocity $\beta\approx 10^{-3}$. Therefore, these experiments do not constrain significantly the coupling $c_a$ in Eq.~(\ref{elastic}), whose contribution to the cross section is suppressed by a factor of $\beta^2$, nor $c_N$, that couples $\chi$ only to neutrinos. In contrast, the coupling $c_s$ implies an unsuppressed spin-independent cross section that must respect the XENON1T bounds \cite{XENON:2018voc}. For $m_\chi$ between 10 GeV and 10 TeV we fit these bounds to
\beq
\sigma_{\chi \N}^{\rm SI}\lsim 0.9\times 10^{-48} \, m_{\chi}^{1 + 169/ m_{\chi}^2}\; {\rm cm}^2,
\eeq
with $m_\chi$ in GeV. This implies 
\beq
{c_s\over \Lambda}\le {c^{\rm max}_s\over \Lambda}\approx  2.5\times 10^{-6} \, { 0.94+m_\chi \over  \sqrt{ m_{\chi}^{1 - 169/ m_{\chi}^2} } }\; {\rm GeV}^{-1}.
\eeq

On the other hand, in $\sigma_{\rm ann}$ (see Eq.~(\ref{ann})) it is the contribution of $c_s$ the one suppressed by a factor of 
$\beta^2$,  with $\beta \approx 1/20$ during freeze out, and $c_a$ and/or $c_N$ should provide the dominant
contribution.  It is straightforward to find the values of these couplings implying a relic abundance $\Omega_c h^2\approx 0.12$ for each value of $m_\chi$. We distinguish the two cases that we plot in Fig.~\ref{f3}: 
\begin{figure}[]
\begin{center}
\includegraphics[scale=0.42]{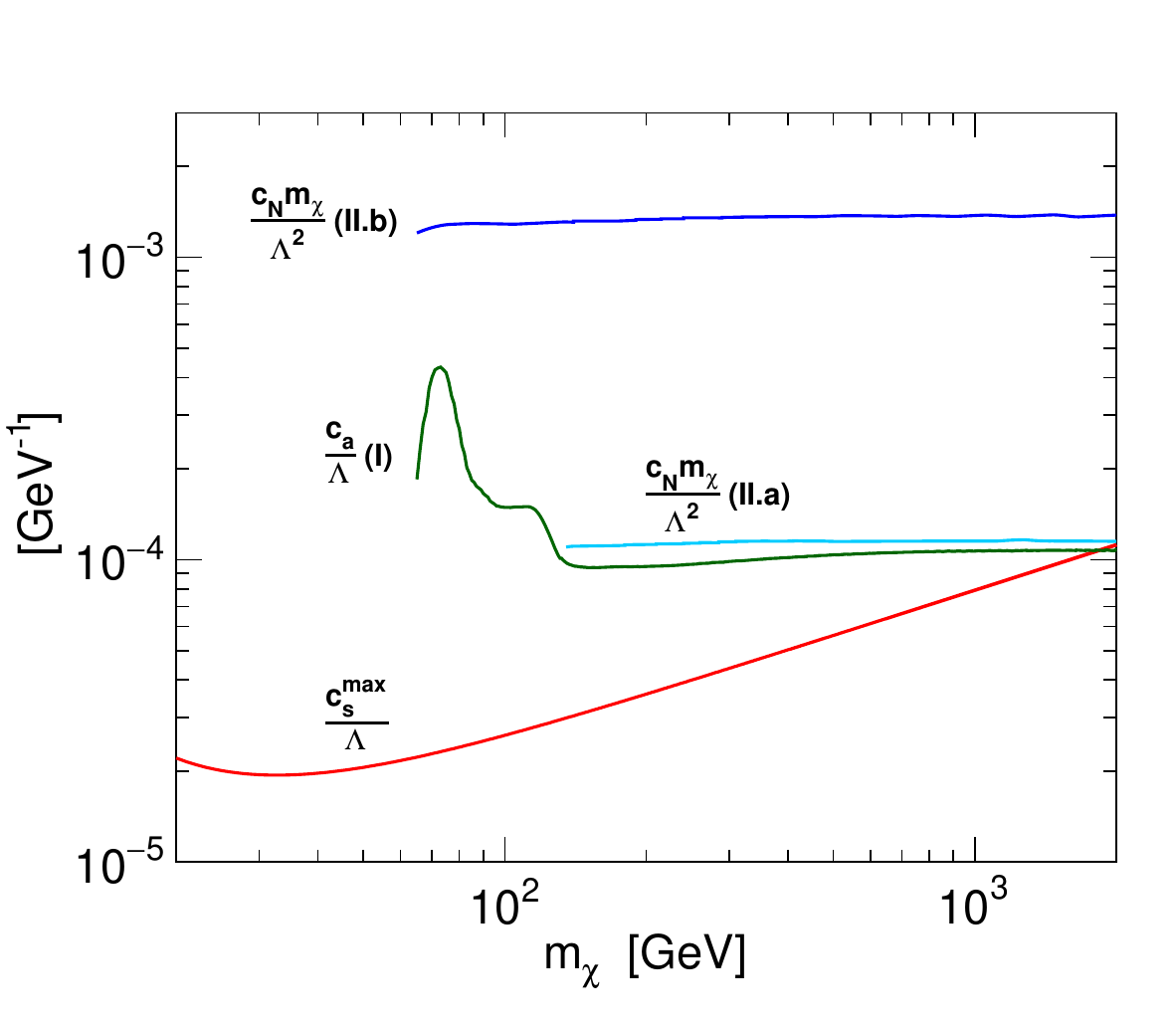}
\end{center}
\vspace{-0.5cm}
\caption{Maximum coupling $c_s/\Lambda$ allowed by direct searches together with the value of $c_a/\Lambda$ (case I) or 
$c_N m_\chi /\Lambda^2$ with $m_N=0.7\,m_\chi$ (case II.a)  and $1.4\,m_\chi$ (II.b) implying $\Omega_c h^2= 0.12$.
\label{f3}}
\end{figure}

\begin{enumerate}
\item [I.] If $c_N\to 0$ there is a value of $c_a/\Lambda$ giving $\Omega_c h^2\approx 0.12$. Notice that the 
channel $\chi \chi \to N \nu$ is in this case subleading and thus the dependence of $c_a/\Lambda$ 
on $m_N$ can be neglected (in the plot we use 
$m_N=1.4\, m_\chi$). 

\item [II.] If $c_a\to 0$ there is also a value of $c_N/\Lambda^2$ giving the right relic abundance. Here, however, we have to distinguish two possibilities: $m_N<m_\chi$, so that the  
channel $\chi \bar \chi \to N \bar N$ is open (we take
$m_N=0.7 m_\chi$), or $m_N>m_\chi$ and a DM particle that annihilates predominantly into $N\nu$, as the contribution to 
$\sigma_{\rm ann}$ of the rest of channels is proportional to $c_s^2\beta^2$.
\end{enumerate}
The general model will combine both cases with weights $\omega_a=c_a^2/\tilde c_a^2$ 
and $\omega_N=c_N^2/\tilde c_N^2$, 
where $\tilde c_{a}$ and $\tilde c_{N}$ are the values of the couplings in the two limiting cases.

\section{Indirect searches at neutrino telescopes} 

We will finally consider the possible signal implied by this model at neutrino telescopes. In particular, we will study whether the signal from DM annihilation in the Sun may be above the background produced by CRs showering in the solar surface \cite{Seckel:1991ffa,Edsjo:2017kjk,Masip:2017gvw,Gutierrez:2019fna,Gutierrez:2022mor} (in the appendix we provide an approximate fit for this CR background). As the capture rate $C=\dd N_\chi^{\rm cap} / \dd t$ of DM by the Sun depends on the same elastic cross section probed at direct searches, we will consider the maximum coupling $c_s^{\rm max}/\Lambda$ consistent with the bounds from XENON1T. 

In our estimate of $C$ we will take the spin-independent DM-nucleon cross section in Eq.~(\ref{ann}),  neglecting the velocity-dependent contribution proportional to $\beta^2\approx 10^{-6}$. 
To deduce the elastic cross section with the different solar nuclei we use the nuclear response functions in \cite{Catena:2015uha}. In particular, we include the collisions with the 6 most abundant nuclei in the Sun (H, He, N, O, Ne, Fe).
We will assume the AGSS09 solar model \cite{Asplund:2009fu}
and the SHM$^{++}$  velocity distribution of the galactic DM \cite{Evans:2018bqy}. 
Our calculation also includes the thermal velocity of the solar nuclei, although its net effect is not important ({\it e.g.}, at $m_\chi=100$ GeV it increases a $5\%$ the capture rate by solar hydrogen but reduces in a 40\% the one by iron, and both effects cancel).
For $m_\chi\ge 10$ GeV and a maximum coupling $c_s^{\rm max}/\Lambda$ we obtain a capture rate that can be fit to
\beq
C^{\rm max}\approx 2.30\times 10^{21} \;m_\chi^{-1-{22\over m_\chi}+{240\over m_{\chi}^2}}\; {\rm s}^{-1},
\eeq
with $m_\chi$ expressed in GeV.

 \begin{figure}[]
\begin{center}
\includegraphics[scale=0.42]{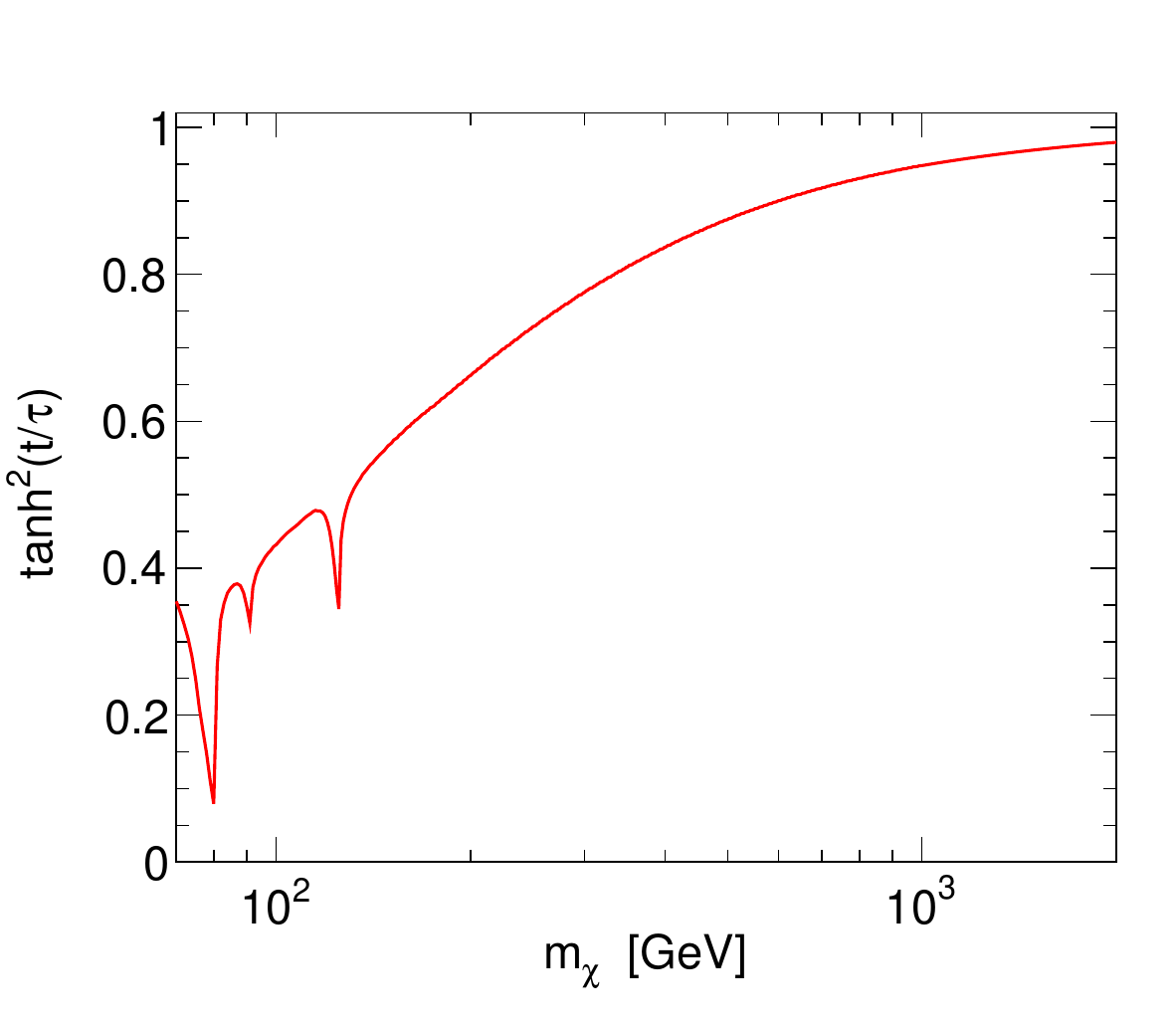}
\end{center}
\vspace{-0.5cm}
\caption{ Suppression  factor $\tanh^2({t_\odot/ \tau})$ for different values of $m_\chi$.
\label{f4}}
\end{figure}
Although in our model the annihilation cross section required to reproduce the relic abundance is relatively large, we need to check whether the age $t_\odot\approx 4.5$ Gyr of the Sun  is long enough to achieve the stationary state where the annihilation rate $\Gamma_A$ is half the capture rate, $\Gamma_A=C/2$. In particular, if we express 
$\Gamma_A\equiv {1\over 2}C_A N_\chi^2$ \cite{Griest:1986yu} 
the time scale $\tau$ for capture and annihilation to equilibrate is just
$\tau=(CC_A)^{-1/2}$, and \cite{Jungman:1995df} 
\beq
{t_\odot\over \tau}\approx 330 \left( C\over {\rm s}^{-1}\right)^{1/2}
\left( { \langle \sigma_{\rm ann} \beta\rangle \over {\rm cm}^3 \,{\rm s}^{-1}} \right)^{1/2}
 \left( m_\chi\over 10\,{\rm GeV} \right)^{3/4}.
\eeq
If $t_\odot/ \tau\gg1$ then $\Gamma_A\approx C/2$, otherwise the annihilation rate  is 
suppressed to 
\beq
\Gamma_A= {C\over 2} \, \tanh^2{t_\odot\over \tau}\,.
\eeq
In Fig.~\ref{f4} we show that, even for the maximum capture rate allowed by bounds from direct searches, this suppression 
cannot be ignored, specially for low values of $m_X$.

\begin{figure}[]
\begin{center}
\includegraphics[scale=0.42]{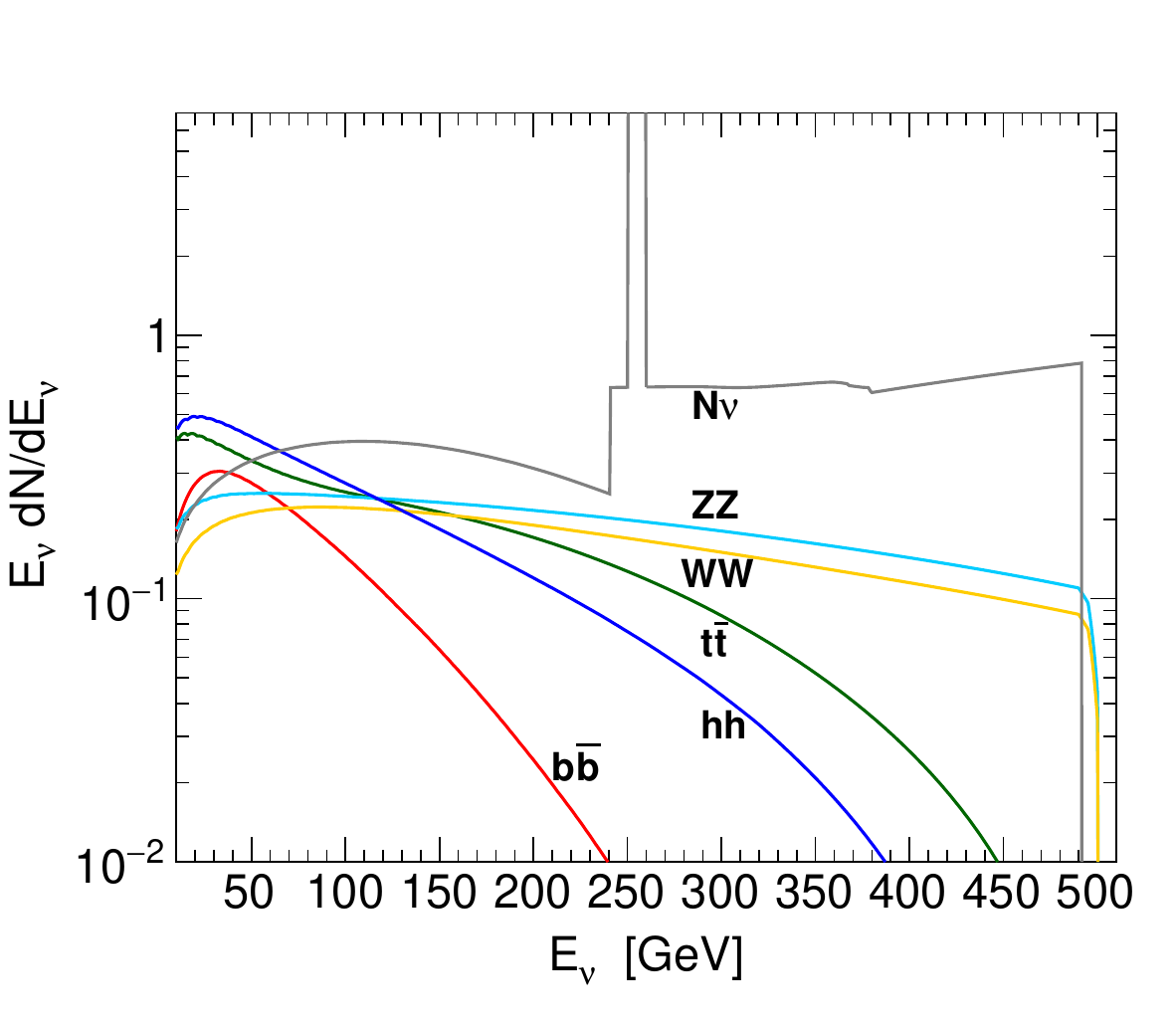}
\end{center}
\vspace{-0.5cm}
\caption{Neutrino yield per one annihilation through each  channel  for $m_\chi= 500$ GeV, $m_N=700$ GeV.
\label{f5}}
\end{figure}

As for the neutrino yields after propagation  from the Sun to the Earth, we use 
the Monte Carlo simulator CHARON \cite{charon}.
In Fig.~\ref{f5} we plot the total neutrino yield ($\nu$ and $\bar \nu$ of all flavors) per annihilation through each channel for $m_\chi=500$ GeV and $m_N=1.4\,m_\chi$. Let us illustrate our results by discussing in some detail a couple of cases.

\subsection{$m_\chi= 300$ GeV, $m_N=105$ GeV}

We have imposed $m_N>m_Z$ to reduce the decays of weak bosons into the 
heavy neutrino ($Z\to N \nu$, $W\to N \ell$, being $\ell$ most frequently the $\tau$ lepton), but we will consider the possibility that $N$ is produced in Higgs decays ($h\to N \nu$) as long as the branching ratio is below 10\%. At the LHC an event with the Higgs giving a heavy neutrino that decays $N\to \ell W$ would appear as a small correction to $h\to W^*W$ plus $W^*\to \nu \ell$ (a process still unobserved for $\ell=\tau$) \cite{ATLAS:2022ooq,ATLAS:2023pwa}, whereas the 10\% increase in the total Higgs width introduced by the new channel is not experimentally excluded. An analogous argument would apply to $h\to N \nu$ with $N\to Z \nu$, which would correct 
$h\to Z^*Z$ plus one (or both \cite{ATLAS:2023tkt})
of the $Z$ bosons decaying into neutrinos \cite{ATLAS:2023ild}.

Therefore, we consider the case with $m_\chi= 300$ GeV, $m_N=105$ GeV, and a Yukawa $\tilde y_N=0.11$ implying BR$(h\to N\nu)=0.1$. We set the maximum coupling consistent with direct bounds, $c^{\rm max}_s/\Lambda = 4.37\times 10^{-5}$ GeV$^{-1}$, and we choose for $c_a/ \Lambda$ and $c_N/ \Lambda^2$ the values giving $\omega_a=0.5=\omega_N$, {\it i.e.}, DM annihilates with equal frequency to $(N\bar N, N\bar \nu, \bar N\nu)$ or through the channels in Fig.~\ref{f2}-left.

At this value of $m_\chi$ the dominant annihilation channel is into Higgs boson pairs. The neutrinos $\nu$ from $\chi \chi \to h h$ plus $h\to N\nu$ are not monochromatic, their energy is distributed between 
\beq
E_\nu^{\rm min}= {1\over 2} \left( m_\chi - \sqrt{m_\chi^2 - m_h^2} - {m_N^2\over m_\chi + \sqrt{m_\chi^2-m_h^2}} \right)
\eeq
and 
\beq
E_\nu^{\rm max}= {1\over 2} \left( m_\chi + \sqrt{m_\chi^2 - m_h^2} - {m_N^2\over m_\chi - \sqrt{m_\chi^2-m_h^2}} \right)\,.
\eeq
However, when $m_\chi$ in near $m_h$ the energy interval shrinks (in the case at hand it goes from
$4$ GeV to $84$ GeV.
The monocromatic channel $\chi \chi \to N\nu$ gives $E_\nu=290$ GeV. 

\begin{figure}[]
\begin{center}
\includegraphics[scale=0.4]{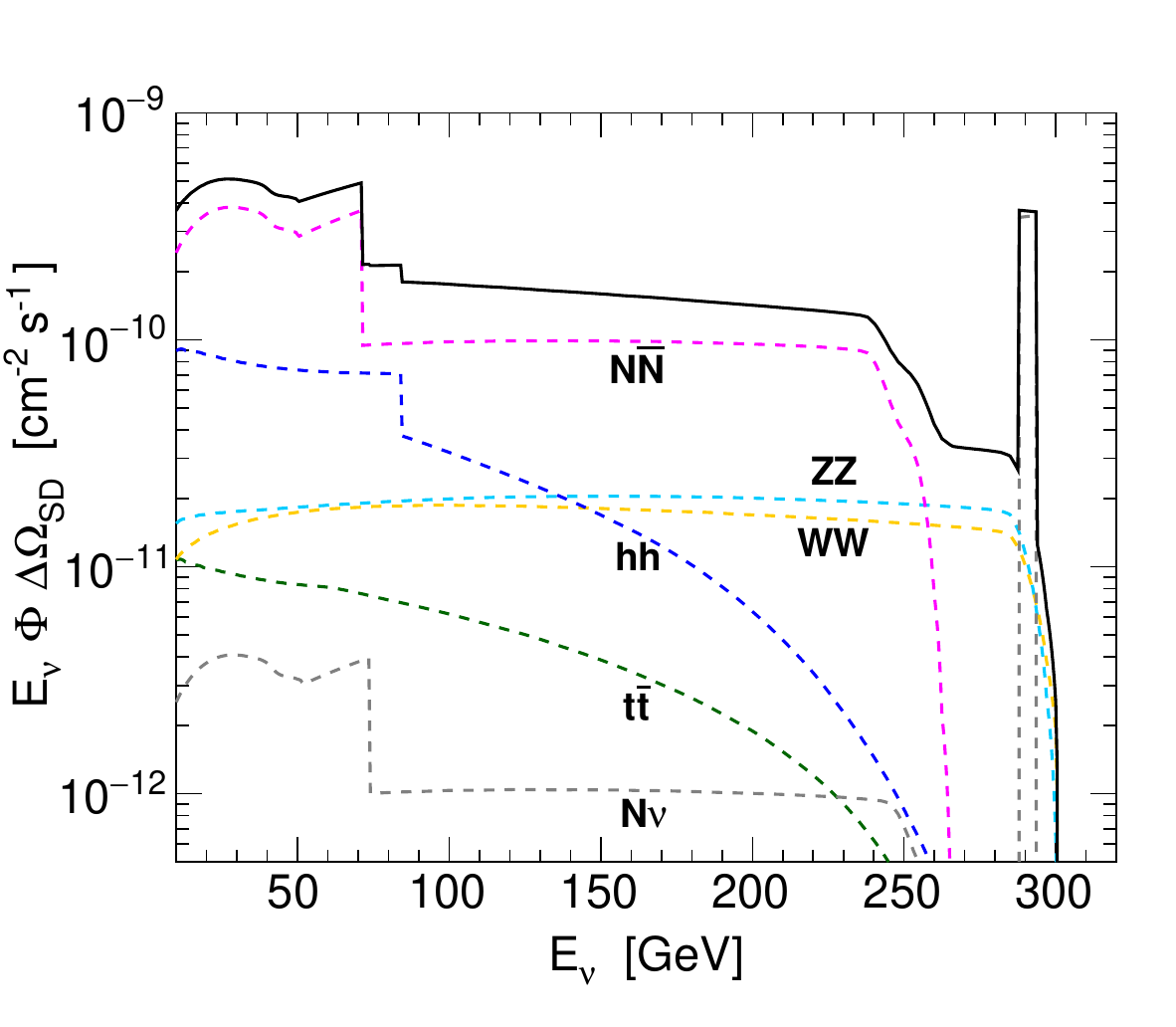}
\includegraphics[scale=0.4]{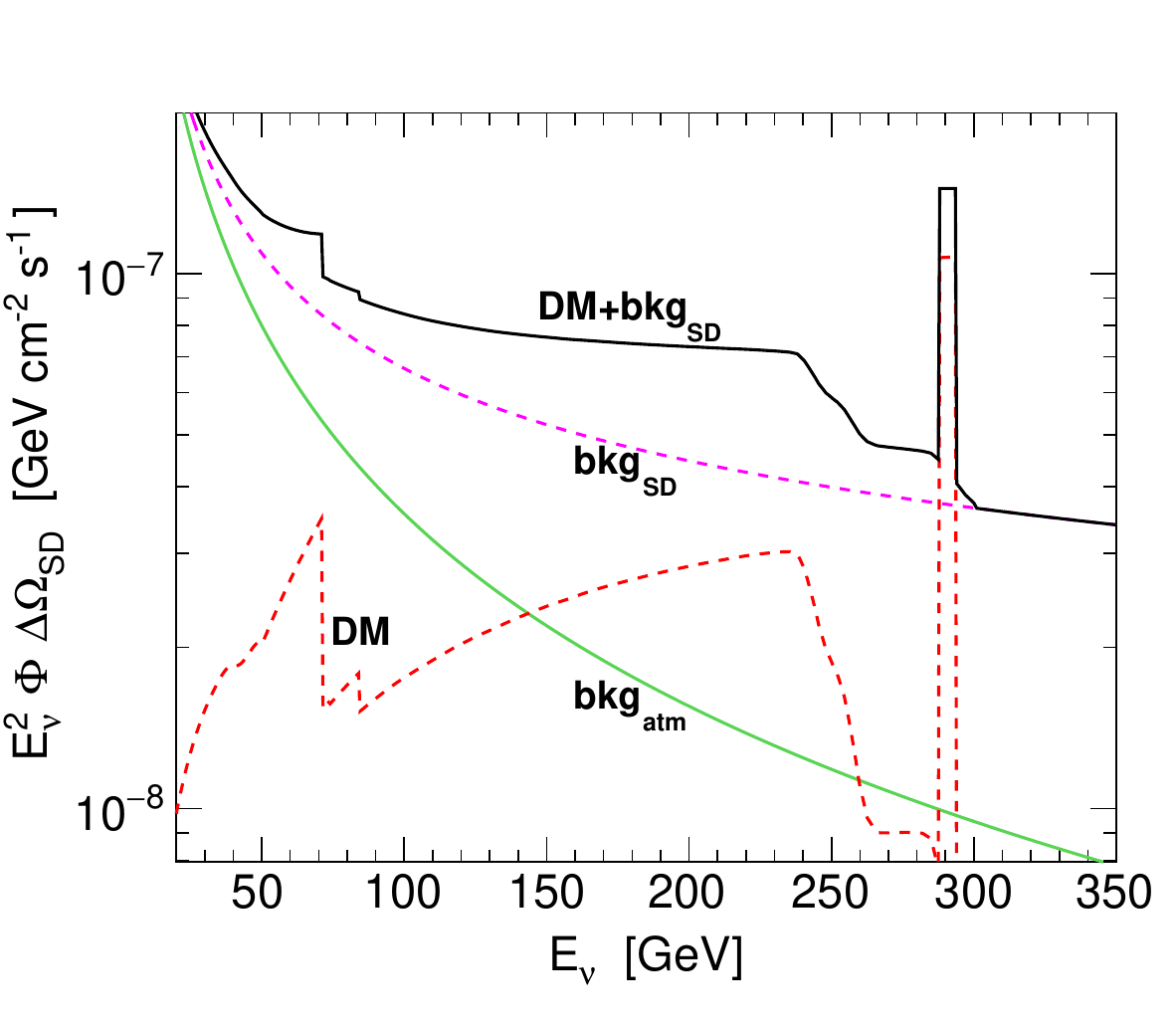}
\end{center}
\vspace{-0.5cm}
\caption{
Solar neutrino flux for $m_\chi= 300$ GeV, $m_N=105$ GeV, $\omega_a=0.5=\omega_N$. On the left, the contribution of the different annihilations channels; on the right, total flux from the SD
including the CR background (for comparison, we include  the atmospheric flux from a {\it fake} Sun). The bin containing the monochromatic $\nu$'s has a width of 6 GeV.
\label{f6}}
\end{figure}

In Fig.~\ref{f6}-left we plot the contribution of each annihilation channel to the neutrino flux from DM annihilation in the Sun.
On the right, we plot the total flux from the solar disk (SD) including both the DM contribution and the neutrino background produced by CR showers in the Sun and by the partial shadow of the Sun (see appendix) for $\theta_z= 45^\circ$. For comparison, we have also included the atmospheric flux from a {\it fake} Sun at the same zenith angle.

\subsection{$m_\chi= 1$ TeV, $m_N=1.4$ TeV}

\begin{figure}[!t]
\begin{center}
\includegraphics[scale=0.4]{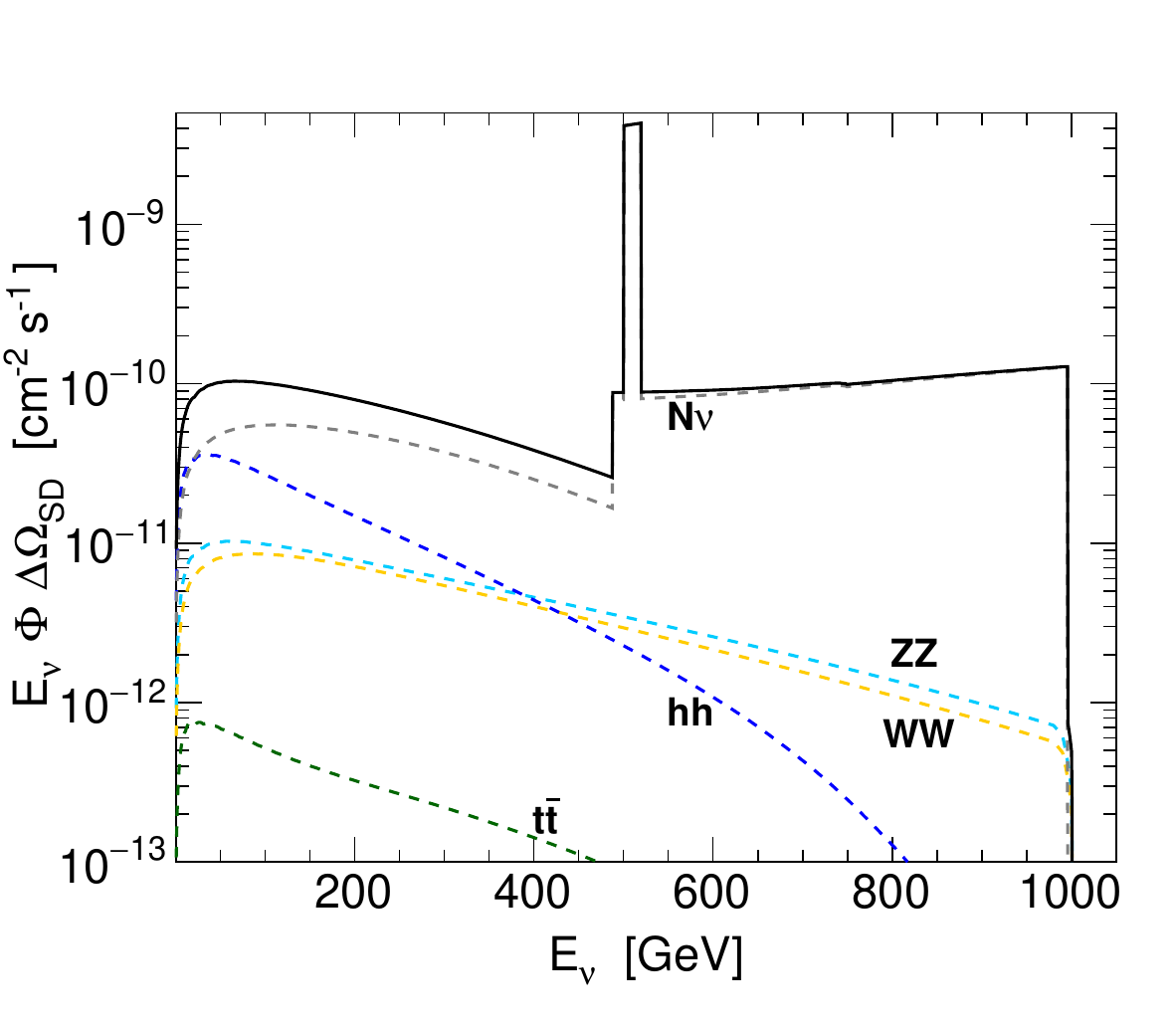}
\includegraphics[scale=0.4]{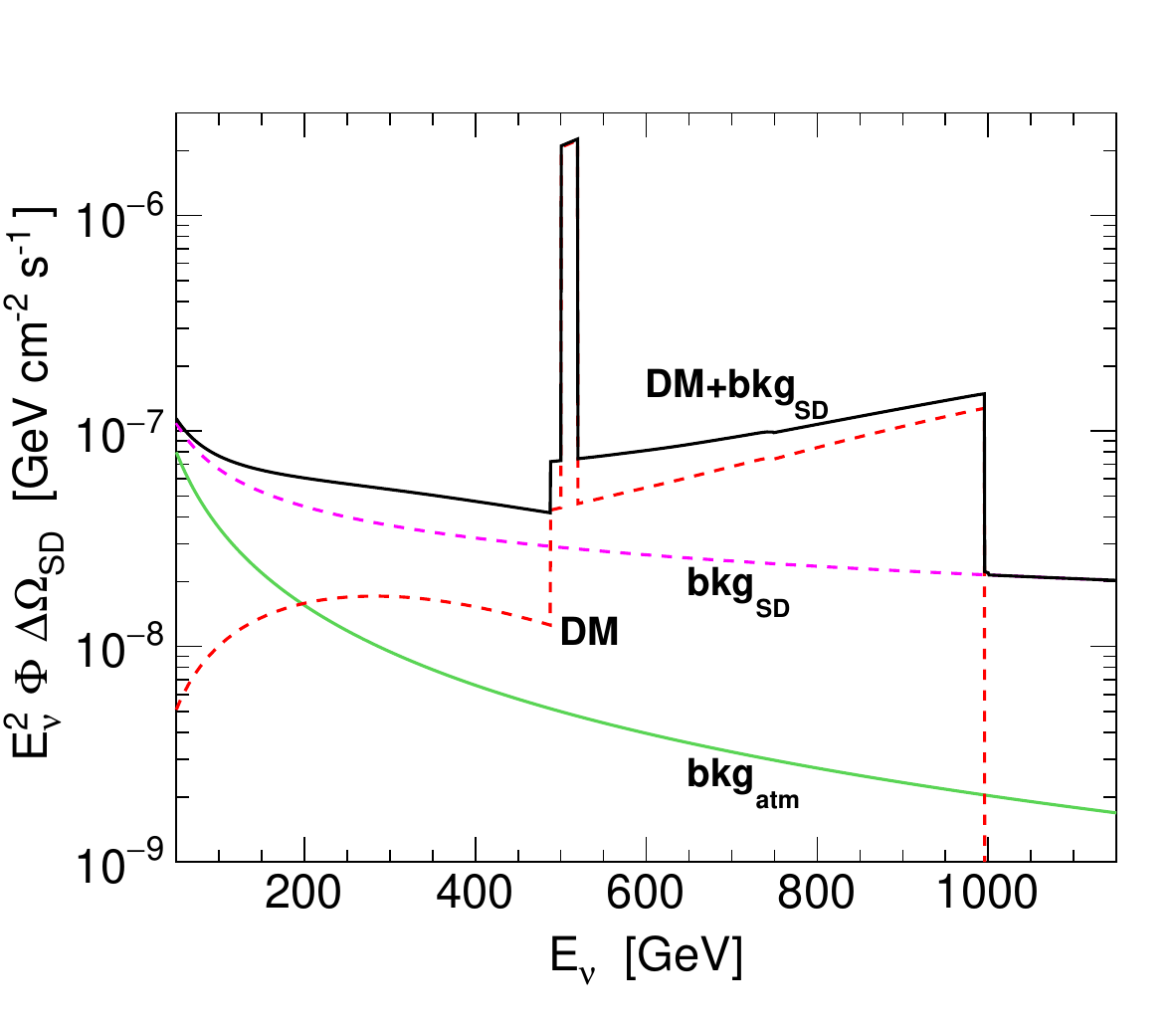}
\end{center}
\vspace{-0.5cm}
\caption{
Neutrino flux from the SD for $m_\chi= 1$ TeV, $m_N=1.4$ TeV, $\omega_a=0.5=\omega_N$. On the left,  contribution of the different annihilations channels; on the right, total flux including the CR background. The bin containing the monochromatic $\nu$'s has a width of 20 GeV.
\label{f7}}
\end{figure}

At these values  of $m_\chi$ and $m_N$ all annihilation channels into standard particles are open but 
$\chi \chi\to N\bar N$ is closed. We take the maximum value of $c_s/\Lambda$ consistent with direct bounds and 
assume that the channels in Fig.~\ref{f2}-left contribute a 50\% to $\sigma_{\rm ann}$, with  $\chi \chi\to N \nu$  providing the rest.
In this case Higgs decays are unaffected by the heavy neutrino, that may only appear in 
flavor analyses \cite{Fernandez-Martinez:2016lgt,Hernandez-Tome:2019lkb,Hernandez-Tome:2020lmh}.

We find a DM signal that is clearly above the CR background from the solar disk: 2.28 neutrinos per km$^{2}$ and second at 500--1000 GeV from DM annihilation (1.62 of them at the 500-520 GeV energy bin), and only 0.25 neutrinos from CRs showers in the solar surface or from the partial CR shadow of the Sun. The atmospheric background from a fake Sun at the same zenith inclination would account for just 0.03 km$^{-2}$ s$^{-1}$ in the same energy interval.

\section{Summary and discussion}  
The WIMP paradigm has been studied for decades, and currently a variety of experiments severely constrains the different models.
In particular, the absence of a signal in direct searches and of missing $E_T$ at the LHC discard the $Z$ and $W$ gauge
bosons as viable mediators of its interactions with the visible sector, as suggested by the so called WIMP miracle.
The Higgs portal appears then as a equally minimal and appealing possibility. Although the Higgs boson has stronger couplings to the heavier SM particles ({\it i.e.}, to itself, the top quark and the weak gauge bosons), it also
admits large Yukawa couplings with the active neutrinos in the presence of heavy Dirac neutral fermions. Those couplings would introduce heavy-light mixings and are actually expected if the origin of neutrino masses is an inverse seesaw mechanism at the TeV scale.

In this work we have analysed such a WIMP scenario: a Majorana fermion interacting through the Higgs portal in a model extended with a heavy Dirac neutrino. 
First we show that, although the WIMP interactions are all spin independent and are thus severely constrained by direct searches, the model naturally implies the observed relic abundance while respecting the current bounds. The question is then whether such a constrained model may provide {\it any} signal in indirect or collider searches at all.

For the neutrino signal from DM annihilation in the Sun, in particular, the same spin independent elastic cross section probed in direct searches also dictates the capture rate by the Sun. In addition, CRs reaching the solar surface produce an irreducible neutrino background that defines a floor in DM searches. 
It has been shown that a capture rate consistent with 
the current XENON1T bounds implies a  neutrino flux from DM annihilations into $\tau^+ \tau^-$, $b \bar b$ or $W^+ W^-$  already below this floor \cite{Gutierrez:2022mor},  {\it i.e.}, these  WIMPs should not give any observable signal there.

The model proposed here, however, is {\it neutrinophilic}. Most important, it includes an annihilation channel
$\chi \chi \to \nu N$ that produces a monochromatic neutrino signal that could be probed at
telescopes \cite{IceCube:2023ies}. The search there for the astrophysical signal from CRs showering in the solar surface may then also reveal this type of DM signal.

The scenario could have interesting implications at the LHC as well \cite{Djouadi:2011aa,CMS:2022dwd}, especially if the heavy neutrino in the model is lighter than the
Higgs boson. The decays $h\to N \nu$ with $N\to  \ell \,W$ (with $\ell$ most frequently the $\tau$ lepton but also the lighter flavors) or $N\to \nu Z$
would appear as a non-standard contribution in $h\to WW^*, ZZ^*$ searches, where 10\% deviations from the SM prediction are not yet experimentally excluded.

In summary, although the WIMP is a DM candidate certainly constrained by several decades of experimental searches,
we think that the presence of heavy Dirac 
neutrinos at the TeV scale revives all the reasons why the scenario is phenomenologically interesting.

\section*{Acknowledgments}
We would like to thank Miki R. Chala, Pablo Olgoso and Jos\'e Santiago for discussions.
This work was partially supported by the Spanish Ministry of Science, Innovation and Universities
(PID2019-107844GB-C21/AEI/10.13039/501100011033) and by the Junta de Andaluc{\'\i}a 
(FQM 101). 

\section*{Appendix A}

Here we provide approximate fits for the atmospheric and solar neutrino fluxes of CR origin that are a background in DM searches. To obtain these fluxes we have followed the procedure in \cite{Gutierrez:2022mor}, 
setting the parameter
$E_c$ to 5 TeV (versus 3--6 TeV in that reference) in order to obtain an optimal fit of recent HAWC data on the 
gamma-ray flux  at 1 TeV \cite{HAWC:2022khj}. In Fig.~\ref{f7} we include that flux together with the model prediction.

\begin{figure}[!t]
\begin{center}
\includegraphics[scale=0.45]{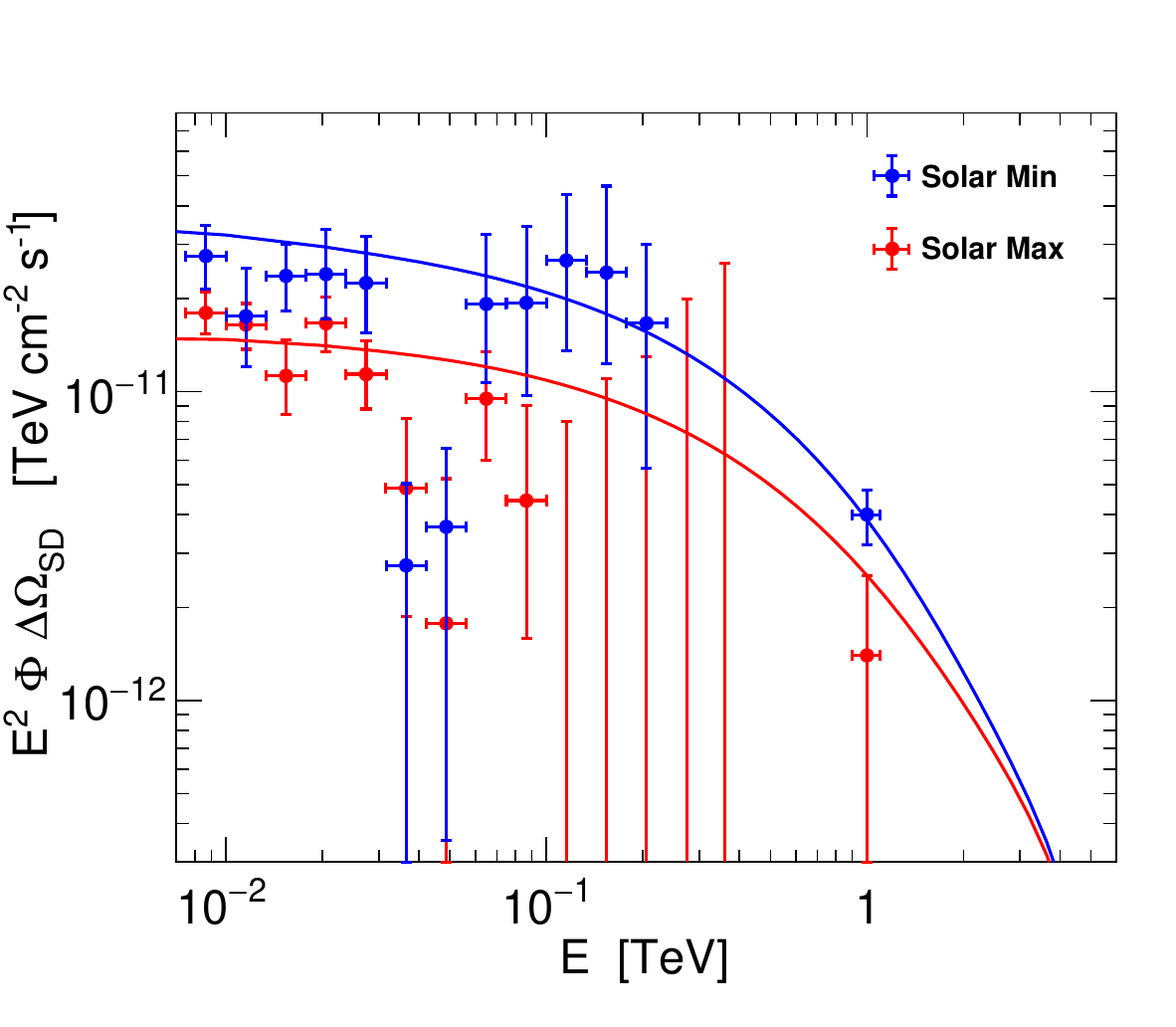}
\end{center}
\vspace{-0.3cm}
\caption{Gamma-ray flux from the solar disk obtained with the model in \cite{Gutierrez:2022mor} for $E_c=5$ TeV and data from HAWC ($E=1$ TeV) \cite{HAWC:2022khj} and
Fermi-LAT (at $E\le 200$ GeV)  \cite{Linden:2018exo}.
\label{f8}}
\end{figure}

In the expressions below we give the neutrino flux integrated over the angular region 
($\Delta\Omega_\odot$) occupied by the Sun, with  $E$ is in GeV, $t$ in years
($t=0$ at the solar minimum), and $\Delta\Omega_\odot \, \Phi_{\nu}$ in GeV$^{-1}$ cm$^{-2}$ s$^{-1}$.
The angle $\theta^*(\theta_z)$ is defined in \cite{Lipari:1993hd, Gutierrez:2021wfy}:
\beq
\tan \theta^* = {R_\oplus \sin\theta_z\over \sqrt{R_\oplus^2 \cos^2\theta_z + \left( 2 R_\oplus + h \right) h }}\,.
\label{theta}
\eeq
For the atmospheric flux we have
\beq
\Delta\Omega_\odot \, \Phi_{\nu_\mu}^{\rm atm} (E,\theta)= 
4.41\times 10^{-6} \;E^{-2.97 - 0.0109 \log E - 
  0.00139 \log^2 E} \,F^{\rm atm}_1(E,\theta)\,;
\eeq
\beq
\Delta\Omega_\odot \, \Phi_{\nu_e}^{\rm atm} (E,\theta)= 
1.94\times 10^{-6}  \;E^{-3.30 - 0.0364 \log^{1.35} E +
0.0103 \log^{1.85} E}\,F^{\rm atm}_2(E,\theta)
\eeq
with
\beq
F^{\rm atm}_1(E,\theta)={\left( {176\over E} \right)^{0.6} + 
\cos [\theta^*({\pi\over 4})] \over \left( {176\over E} \right)^{0.6}  + \cos[\theta^*(\theta_z)]}\,;\;\;
F^{\rm atm}_2(E,\theta)={\left( {7.5\times 10^{-4}\over E} \right)^{0.21} + 
\cos [\theta^*({\pi\over 4})] \over \left( {7.5\times 10^{-4}\over E} \right)^{0.21}  + \cos[\theta^*(\theta_z)]}\,.
\eeq
For the atmospheric neutrinos from the partial cosmic ray shadow of the Sun and from neutrons produced in the
solar surface and  reaching the Earth the flux is
\beq
\Delta\Omega_\odot \, \Phi_{\nu_\mu}^{\rm shad+n}(E,\theta,t) = 
4.33\times 10^{-6} \;E^{G^{\rm atm}_1(E,t)} \,F^{\rm atm}_2(E,\theta)\,;
\eeq
\beq
\Delta\Omega_\odot \, \Phi_{\nu_e}^{\rm shad+n}(E,\theta,t) = 
1.37\times 10^{-6}\;E^{G^{\rm shad+n}_2(E,t)}\,F^{\rm atm}_2(E,\theta)
\eeq
with
\beq
G^{\rm shad+n}_1(E,t)=-2.98 - 0.017 \log E +
 0.012 \cos {2\pi t\over 11} \log^2 E  - 3.3\times 10^{-4} \log^3 E \nonumber 
 \eeq
\vspace{-0.9cm}
\beq
- 3.9\times 10^{-6} \log^5 E\,; \hspace{3.9cm}
\eeq
\beq
G^{\rm shad+n}_2(E,t)=-3.1 - 0.061 \log E - 
\cos {2\pi t\over 11} \left(
0.00305 \log E  +2.1\times 10^{-6} \log^5 E\right)  \nonumber 
\eeq
\vspace{-0.9cm}
\beq
-5.4\times 10^{-7} \log^6 E\,.\hspace{4.2cm}
\eeq
Finally, the neutrinos produced by cosmic rays in the Sun reach us in the three flavors with the same frequency and a flux per
flavor
\beq
\Delta\Omega_\odot \, \Phi_{\nu_i}^{\rm \odot} (E,t)= 6.0 \times 10^{-9}
\left( 1 - {405 \sin^2 {\pi t\over 11}\over 900+E}  \right)
E^{-1.20 - 0.1 \log E - 
  0.0042 \log^2 E+ 1.6\times 10^{-5} \log^4 E} \,.
\eeq

\end{document}